\documentclass[preprint,12pt, a4paper]{elsarticle}

\usepackage[utf8]{inputenc}
\usepackage[T1]{fontenc}
\usepackage[english]{babel}
\usepackage{color,soul}
\usepackage[table,dvipsnames]{xcolor}
\usepackage{booktabs}
\usepackage{threeparttable}
\usepackage{amsfonts} 
\usepackage{nicefrac}  
\usepackage{microtype} 
\usepackage{lipsum}
\usepackage{graphicx}
\usepackage{subcaption}
\usepackage[export]{adjustbox}

\usepackage{bbding}

\definecolor{myblue}{RGB}{13, 71, 161} 
\definecolor{mygreen}{RGB}{115, 140, 84} 
\definecolor{myred}{RGB}{216, 28, 56} 

\usepackage{acro}
\DeclareAcroEnding{possessive}{'s}{'s}
\NewAcroCommand\acg{m}{\acropossessive\UseAcroTemplate{first}{#1}}
\NewAcroCommand\acsg{m}{\acropossessive\UseAcroTemplate{short}{#1}}
\NewAcroCommand\aclg{m}{\acropossessive\UseAcroTemplate{long}{#1}}
\NewAcroCommand\acfg{m}{%
    \acrofull
    \acropossessive
    \UseAcroTemplate{first}{#1}%
}
\NewAcroCommand\iacsg{m}{%
    \acroindefinite
    \acropossessive
    \UseAcroTemplate{short}{#1}%
}
\DeclareAcronym{AGN}{
    short = AGN,
    long  = Active Galactic Nuclei
}
\DeclareAcronym{AK}{
    short = AK,
    long  = analytic kludge
}
\DeclareAcronym{AAK}{
    short = AAK,
    long  = argumented analytic kludge
}
\DeclareAcronym{AUC}{
    short = AUC,
    long  = area under the curve
}
\DeclareAcronym{AI}{
    short = AI,
    long = artificial intelligence
}
\DeclareAcronym{BH}{
    short = BH,
    long  = black hole
}
\DeclareAcronym{BHB}{
    short = BHB,
    long  = black hole binary,
    long-plural-form = black hole binaries
}
\DeclareAcronym{BBH}{
    short = BBH,
    long  = binary black hole
}
\DeclareAcronym{CNN}{
    short = CNN,
    long  = convolutional neural network
}
\DeclareAcronym{CV}{
    short = CV,
    long  = computer vision
}

\DeclareAcronym{DECIGO}{
    short = DECIGO,
    long  = DECi-hertz Interferometer Gravitational wave Observatory
}
\DeclareAcronym{DECODE}{
    short = DECODE,
    long  = DilatEd COnvolutional neural network for Detecting Extreme-mass-ratio inspirals
}

\DeclareAcronym{DNN}{
    short = DNN,
    long  = deep neural network
}

\DeclareAcronym{DL}{
    short = DL,
    long  = deep learning
}
\DeclareAcronym{ESA}{
    short = ESA,
    long  = European Space Agency
}
\DeclareAcronym{EMRI}{
    short = EMRI,
    long  = extreme-mass-ratio inspiral
}
\DeclareAcronym{FPR}{
    short = FPR,
    long  = false positive rate
}
\DeclareAcronym{GB}{
    short = GB,
    long  = galactic binary,
    long-plural-form = galactic binaries
}
\DeclareAcronym{GR}{
    short = GR,
    long  = general relativity
}
\DeclareAcronym{GW}{
    short = GW,
    long  = gravitational wave
}
\DeclareAcronym{GWDA}{
    short = GWDA,
    long  = gravitational wave data analysis
}
\DeclareAcronym{LDC}{
    short = LDC,
    long  = LISA Data Challenge
}
\DeclareAcronym{LIGO}{
    short = LIGO,
    long  = \href{http://www.ligo.caltech.edu/}{Laser Interferemeter Gravitational Wave Observatory}
}
\DeclareAcronym{LISA}{
    short = LISA,
    long  = \href{https://www.lisamission.org/}{Laser Interferometer Space Antenna}
}

\DeclareAcronym{LSO}{
    short = LSO,
    long  = last stable orbit
}
\DeclareAcronym{MBH}{
    short = MBH,
    long  = massive black hole
}
\DeclareAcronym{MBHB}{
    short = MBHB,
    long  = massive black hole binary,
    long-plural-form = massive black hole binaries
}

\DeclareAcronym{MCMC}{
    short = MCMC,
    long  = Markov-chain Monte Carlo
}
\DeclareAcronym{MLDC}{
    short = MLDC,
    long  = \href{http://astrogravs.nasa.gov/docs/mldc/}{Mock LISA Data Challenge}
}
\DeclareAcronym{MLP}{
    short = MLP,
    long  = multi-layer perceptron
}
\DeclareAcronym{NK}{
    short = NK,
    long  = numerical kludge
}

\DeclareAcronym{NLP}{
    short = NLP,
    long  = natural language processing
}
\DeclareAcronym{OMS}{
    short = OMS,
    long  = optical metrology system
}
\DeclareAcronym{PSD}{
    short = PSD,
    long  = power spectral density
}
\DeclareAcronym{ReLU}{
    short = ReLU,
    long  = rectified linear 
 unit
}

\DeclareAcronym{ROC}{
    short = ROC,
    long  = receiver operating characteristic
}
\DeclareAcronym{SGWB}{
    short = SGWB,
    long  = stochastic gravitational wave background
}
\DeclareAcronym{SMBH}{
    short = SMBH,
    long  = super-massive black hole
}
\DeclareAcronym{SNR}{
    short = SNR,
    long  = signal-to-noise ratio
}
\DeclareAcronym{SOBH}{
    short = SOBH,
    long  = stellar origin black hole binary
}
\DeclareAcronym{SSB}{
    short = SSB,
    long  = solar system barycenter
}
\DeclareAcronym{TCN}{
    short = TCN,
    long  = temporal convolutional network
}
\DeclareAcronym{TDI}{
    short = TDI,
    long  = time delay interferometry
}
\DeclareAcronym{TPR}{
    short = TPR,
    long  = true positive rate
}
\DeclareAcronym{t-SNE}{
    short = t-SNE,
    long  = t-distributed stochastic neighbor embedding
}
\DeclareAcronym{VGB}{
    short = VGB,
    long  = verification galactic binary,
    long-plural-form = verification galactic binaries
}

\graphicspath{{Fig/}}

\usepackage{amssymb}
\usepackage{hyperref}

\journal{SoftwareX}

\begin{document}
\renewcommand{\labelenumii}{\arabic{enumi}.\arabic{enumii}}

\begin{frontmatter}
	% \title{GWAI: Harnessing Artificial Intelligence for Enhanced Gravitational Wave Data Analysis}
	\title{GWAI: Artificial Intelligence Platform for Enhanced Gravitational Wave Data Analysis}

	\author[1,2,3,4]{Tianyu Zhao}
	\author[2]{Yue Zhou}
	\author[3,4]{Ruijun Shi}
	\author[3,4,5]{Zhoujian Cao\corref{cor1}}
	\ead{zjcao@bnu.edu.cn}
	\author[2]{Zhixiang Ren\corref{cor1}}
	\ead{renzhx@pcl.ac.cn}

	\address[1]{Center for Gravitational Wave Experiment, National Microgravity Laboratory, Institute of Mechanics, Chinese Academy of Sciences, Beijing, 100190, China}
	\address[2]{Peng Cheng Laboratory, Shenzhen, China, 518055}
	\address[3]{School of Physics and Astronomy, Beijing Normal University, Beijing, China, 100875}
	\address[4]{Institute for Frontiers in Astronomy and Astrophysics, Beijing Normal University, Beijing, China, 102206}
	\address[5]{School of Fundamental Physics and Mathematical Sciences, Hangzhou Institute for Advanced Study, UCAS, Hangzhou, China, 310024}

	\cortext[cor1]{Corresponding author}

	\begin{abstract}
		Gravitational wave (GW) astronomy has opened new frontiers in understanding the cosmos, while the integration of artificial intelligence (AI) in science promises to revolutionize data analysis methodologies.
		However, a significant gap exists, as there is currently no dedicated platform that enables scientists to develop, test, and evaluate AI algorithms efficiently for GW data analysis.
		To address this gap, we introduce GWAI, a pioneering AI-centered software platform designed for GW data analysis. GWAI contains a three-layered architecture that emphasizes simplicity, modularity, and flexibility, covering the entire analysis pipeline.
		GWAI aims to accelerate scientific discoveries, bridging the gap between advanced AI techniques and astrophysical research.
	\end{abstract}


\end{frontmatter}

%\linenumbers

\section*{Metadata}
\label{sec:meta}

\begin{table}[!ht]
	\begin{tabular}{|l|p{6.5cm}|p{6.5cm}|}
		\hline
		C1 & Current code version                                             & v1.0                                                                                                                 \\
		\hline
		C2 & Permanent link to code/repository used for this code version     & \url{https://github.com/AI-HPC-Research-Team/GWAI}                                                                   \\
		\hline
		C3 & Permanent link to Reproducible Capsule                           & N/A                                                                                                                  \\
		\hline
		C4 & Legal Code License                                               & GNU General Public License (GPL)                                                                                     \\
		\hline
		C5 & Code versioning system used                                      & git                                                                                                                  \\
		\hline
		C6 & Software code languages, tools, and services used                & \texttt{Python 3.8}, \texttt{CUDA 11.4}, \texttt{fftw 3.3.10-1}, \texttt{gsl 2.7.1}                                  \\
		\hline
		C7 & Compilation requirements, operating environments \& dependencies & \texttt{Ubuntu 20.04}, \texttt{gcc 9.4.0}, \texttt{cmake 3.28.2}                                                     \\
		\hline
		C8 & If available Link to developer documentation/manual              & \url{https://gwai.readthedocs.io/en/latest/index.html}                                                               \\
		\hline
		C9 & Support email for questions                                      & \href{mailto:renzhx@pcl.ac.cn}{\texttt{renzhx@pcl.ac.cn}}, \href{mailto:zjcao@bnu.edu.cn}{\texttt{zjcao@bnu.edu.cn}} \\
		\hline
	\end{tabular}
	\caption{Code metadata}
	\label{codeMetadata}
\end{table}

\section{Motivation and significance}
% 1. GW Detection
The direct detection of \acp{GW}, as predicted by Einstein's \ac{GR}, was a seminal event in astrophysics, first achieved in 2015 \cite{abbott_observation_2016}. These disturbances in spacetime, emanating from cataclysmic astronomical occurrences like black hole collisions, have inaugurated an unprecedented era in astronomical observation \cite{the_ligo_scientific_collaboration_gwtc-3_2021}. This discovery not only substantiated a crucial aspect of \ac{GR} but also provided a new window for probing the universe's most energetic events \cite{bailes_gravitational-wave_2021}.

% GWDA
Transitioning from the initial detection of \acp{GW} to the intricate discipline of \acl{GWDA} represents a paradigm shift in astrophysical research \cite{usman_pycbc_2016}. The field of \acl{GWDA} is fraught with challenges, including the extraction of extremely weak signals from overwelming detector noise and the management of complex, voluminous datasets \cite{finn_detection_1992}. This evolution underscores a shift from mere observation to a sophisticated analysis of \acp{GW}, aiming to unveil the intricacies of astronomical phenomena like black holes and neutron stars \cite{bailes_gravitational-wave_2021}.

% 2. Space-based GWDA (Taiji)
Space-based \ac{GW} detection projects, like the \ac{LISA} \cite{amaro-seoane_laser_2017}, Taiji \cite{hu_taiji_2017,ren_taiji_2023}, and TianQin \cite{luo_tianqin_2016} program, represent pivotal advancements in this field. These systems operate far from the Earth's noise and disturbances and aim to detect \acp{GW} with unprecedented precision \cite{babak_science_2017}. The Taiji program is an ambitious project by the Chinese Academy of Science that is set to complement and extend the capabilities of earth-based detectors. It is potentially promising to revolutionize our understanding of \acp{GW} and the very fabric of spacetime \cite{thetaijiscientificcollaboration_taiji_2021}.

% 1. traditional pipelines
The landscape of \acl{GWDA} has been significantly shaped by the development of specialized software tools, each contributing unique capabilities to the field. PyCBC \cite{usman_pycbc_2016} has emerged as a prominent tool, particularly for analyzing data from \ac{GW} detectors like \ac{LIGO} \cite{the_ligo_scientific_collaboration_2015}, Virgo \cite{acernese_advanced_2015}, and KAGRA \cite{kagra_collaboration_kagra_2019}, using matched filtering techniques to detect signals from astrophysical sources \cite{finn_detection_1992}. GstLAL \cite{cannon_gstlal_2021}, developed within the GStreamer framework, offers robust capabilities for noise suppression and signal extraction and is crucial for parameter estimation in \ac{GW} studies. Additionally, Coherent WaveBurst \cite{drago_coherent_2021} has been instrumental in identifying and characterizing transient gravitational waveforms, employing a time-frequency analysis approach that is adept at handling non-stationary noise in detector data. Another notable tool, BayesWave \cite{cornish_bayeswave_2015,cornish_bayeswave_2021}, utilizes Bayesian statistical methods to differentiate between astrophysical signals and detector glitches and provides a framework for signal reconstruction and parameter estimation. These tools are fundamental in analyzing \ac{GW} data and have enabled a multitude of groundbreaking discoveries \cite{the_ligo_scientific_collaboration_gwtc-3_2021}.
However, as \acl{GWDA} evolves with increasingly complex datasets, there is a growing need for more advanced processing techniques \cite{bambi_advances_2021,yaseen_current_2022}.
% The integration of \ac{AI} and machine learning into \acl{GWDA}, represents a significant advancement, building upon the foundational work of these traditional tools 

% 2. Deep Learning in GWDA
In recent years, the application of \ac{DL} techniques in \acl{GWDA} has gained significant momentum \cite{cuoco_machine_2020,zhao_dawning_2023}. These advanced methodologies have been instrumental across various facets of \acl{GWDA}, including signal detection \cite{gabbard_matching_2018}, glitch classification \cite{razzano_image-based_2018}, denoising \cite{chatterjee_extraction_2021-3}, waveform generation \cite{khan_interpretable_2022}, and parameter estimation \cite{dax_flow_2023}. The potency of \ac{DL} in this context lies in its ability to discern subtle patterns in complex data.
% a crucial ability given the intricate nature of \ac{GW} signals. 
Several studies have successfully employed deep neural networks to enhance the accuracy and efficiency of \acl{GWDA} tasks, thereby contributing to a deeper understanding of \ac{GW} phenomena (see \cite{zhao_dawning_2023} and references therein). These advancements in applying \ac{DL} to \acl{GWDA} have marked a transformative shift in how \ac{GW} data is processed and analyzed \cite{huerta_accelerated_2021}.

% 3. Need for an Open Source GWDA Platform
Despite the growing use of \ac{DL} in \acl{GWDA}, there remains a notable gap in the availability of open-source software platforms dedicated to this field. Many papers and studies do not provide accessible open source code, which is in stark contrast to other research fields like \ac{CV} \cite{dosovitskiy_image_2020}, \ac{NLP} \cite{devlin_bert_2019} and time-series analysis \cite{nie_time_2022, subakan_attention_2021},
% and speech recognition \cite{subakan_attention_2021}, 
where the availability of open source libraries has been instrumental in driving research success \cite{tsai,ravanelli_speechbrain_2021}.
% These fields have demonstrated the substantial benefits of having readily available, community-driven software tools, which have been widely adopted and have significantly accelerated research progress. 
Therefore, it is crucial to address this gap in \acl{GWDA} by developing a comprehensive, open-source software platform. Such a platform would not only facilitate the broader application and validation of \ac{DL} techniques in \acl{GWDA} but also foster a collaborative research environment, accelerating innovations and discoveries in the field.

% original fig. 1 position 

% 4. In this paper:
In this paper, we present a pioneering development in \acl{GWDA}.
% : the first AI-based software platform tailored for this field. 
The key features of our platform are:
\begin{itemize}
	\item \textbf{First AI-Centred GW Platform:} We introduce the first AI-driven data analysis software in \acl{GWDA}, setting a new standard in the field with advanced algorithms and intelligent data processing capabilities.
	\item \textbf{Comprehensive Coverage of Applications:} Our platform encompasses the entire \acl{GWDA} pipeline, from initial data acquisition to final analysis, ensuring a thorough and integrated approach to \ac{GW} studies.
	\item \textbf{Modularized and User-Friendly:} With an emphasis on simplicity, modularity, and flexibility, the platform is not only easy to navigate but also adaptable to various research needs. Comprehensive documentation and user guides enhance its accessibility, making it suitable for both seasoned researchers and newcomers to the field.
\end{itemize}
These attributes establish our platform as a toolbox in \acl{GWDA}, and blend advanced \ac{AI} techniques with user-centric design to facilitate cutting-edge \ac{GW} research.

\section{Software description}

\subsection{Software architecture}
\begin{figure}[h!]
	\centering
	\includegraphics[width=0.6\textwidth]{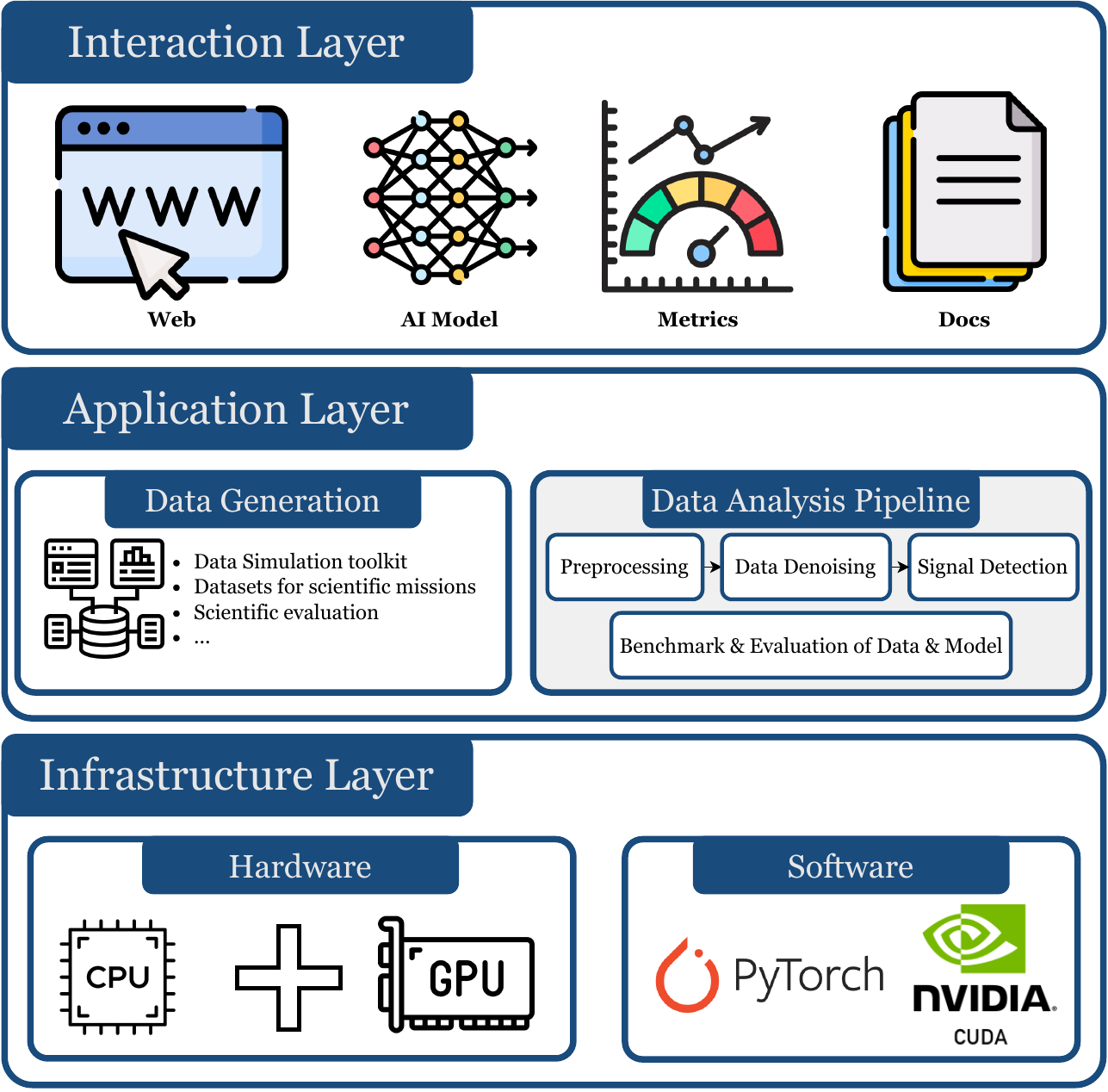}
	\caption{\textbf{Overall architecture of the GWAI platform:}
		%Illustrated here is the multi-layered structure of the GWAI platform, encompassing the infrastructure, application, and interaction layers. 
		The infrastructure layer forms the foundation with its software and hardware resources. At the application layer, the platform is divided into two primary components: the data generation module and the data analysis pipeline, which facilitate comprehensive \ac{GW} data processing. The interaction layer enhances user engagement through a Web UI, provides access to models and functionalities via API, and offers extensive documentation for ease of use.}
	\label{fig:platform}
\end{figure}
% In developing this groundbreaking \acl{GWDA} software platform, we adhered to key design principles that align with the cutting-edge requirements of the field while ensuring user accessibility and ease of implementation. These principles are:

In developing this groundbreaking \acl{GWDA} software platform, we structured it around a robust 3-layer architecture:
% to meet the cutting-edge requirements of the field while ensuring user accessibility and ease of implementation. This architecture comprises three key layers: 
Infrastructure layer, Application layer, and Interaction layer. These layers collectively ensure a comprehensive, modular, and flexible platform designed to facilitate advanced gravitational wave data analysis. The design principles guiding this architecture are:

\paragraph{Simplicity} The core design principle of our platform is simplicity. We strive to create an interface and underlying mechanics that are straightforward and intuitive. This approach allows users, from seasoned researchers to early students in the field, to engage with the platform without a steep learning curve. The simplicity characteristic ensures that the focus remains on the analysis and interpretation of \ac{GW} data rather than on the installation complexities of the software itself.

\paragraph{Modularity} Recognizing the diverse and evolving needs of the \acl{GWDA} community, our platform is built on a modular framework. This allows users to easily customize their analysis pipeline by adding or removing components as needed. Each module, whether data preprocessing, signal detection, or statistical analysis, is wrapped with independent functions that integrate seamlessly with others. This modularity not only caters to a wide range of research applications but also facilitates the incorporation of future advancements in the field.

\paragraph{Flexibility} Flexibility is another cornerstone of our platform. GWAI can be adapted to various research demands and data types encountered in \ac{GW} analysis. The platform's architecture is crafted to handle a range of data formats and sizes, from small-scale experimental data to large-scale observational datasets. This flexibility extends to the integration of new algorithms and methodologies, empowering users to tailor the platform to their specific research goals.

\paragraph{Ease to Use:}
% To maximize the platform's accessibility, we have invested significant effort in creating a rich repository of resources. This includes detailed documentation, example showcases, and tutorial notebooks, all designed to guide users through the platform's features and capabilities. The inclusion of Jupyter notebooks and other interactive resources makes it easier for newcomers to familiarize themselves with both the platform and the broader field of \acl{GWDA}. Moreover, code snippets in documentation and comprehensive tutorial files provide practical insights into using the platform for various \acl{GWDA} tasks.

To maximize the platform's accessibility, we have invested significant effort in creating a rich repository of resources. This includes detailed documentation, example showcases, and tutorial notebooks, all designed to guide users through the platform's features and capabilities.
% The platform features a Web UI for our datasets, allowing users to interact with and explore the data easily. 
Additionally, YAML configuration files facilitate straightforward setup and customization of experiments.
% Interactive Jupyter notebooks and visualization scripts provide practical tools for analyzing data and interpreting results. 
These resources make it easier for newcomers to familiarize themselves with both the platform and the broader field of \acl{GWDA}. Moreover, code snippets in documentation and comprehensive tutorial files offer practical insights into using the platform for various \acl{GWDA} tasks.

The GWAI platform comprises two primary components: data generation and data analysis, as illustrated in Fig. \ref{fig:platform}. Each module within these components is detailed in the subsections that follow.

\subsection{Software functionalities}
\subsubsection{Data}
% Fig. 3 sub-fig of Fig. 2  code
\begin{figure}[ht!]
	\centering
	\includegraphics[width=0.6\textwidth]{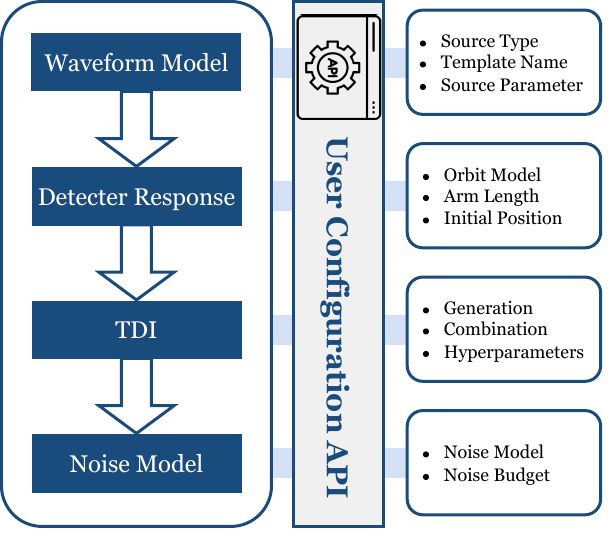}
	\caption{\textbf{Flowchart of the data generation module:} This diagram delineates the sequential process of generating synthetic data, depicted on the left side, from waveform generation through to noise addition, culminating in the synthetic data. On the right, we highlight the module's highly customizable configuration API, designed to tailor the data generation process to specific needs.}
	\label{fig:data}
\end{figure}

\begin{figure}[ht!]
	\centering
	\begin{subfigure}[t]{0.49\textwidth}{
			\includegraphics[width=\textwidth,valign=t]{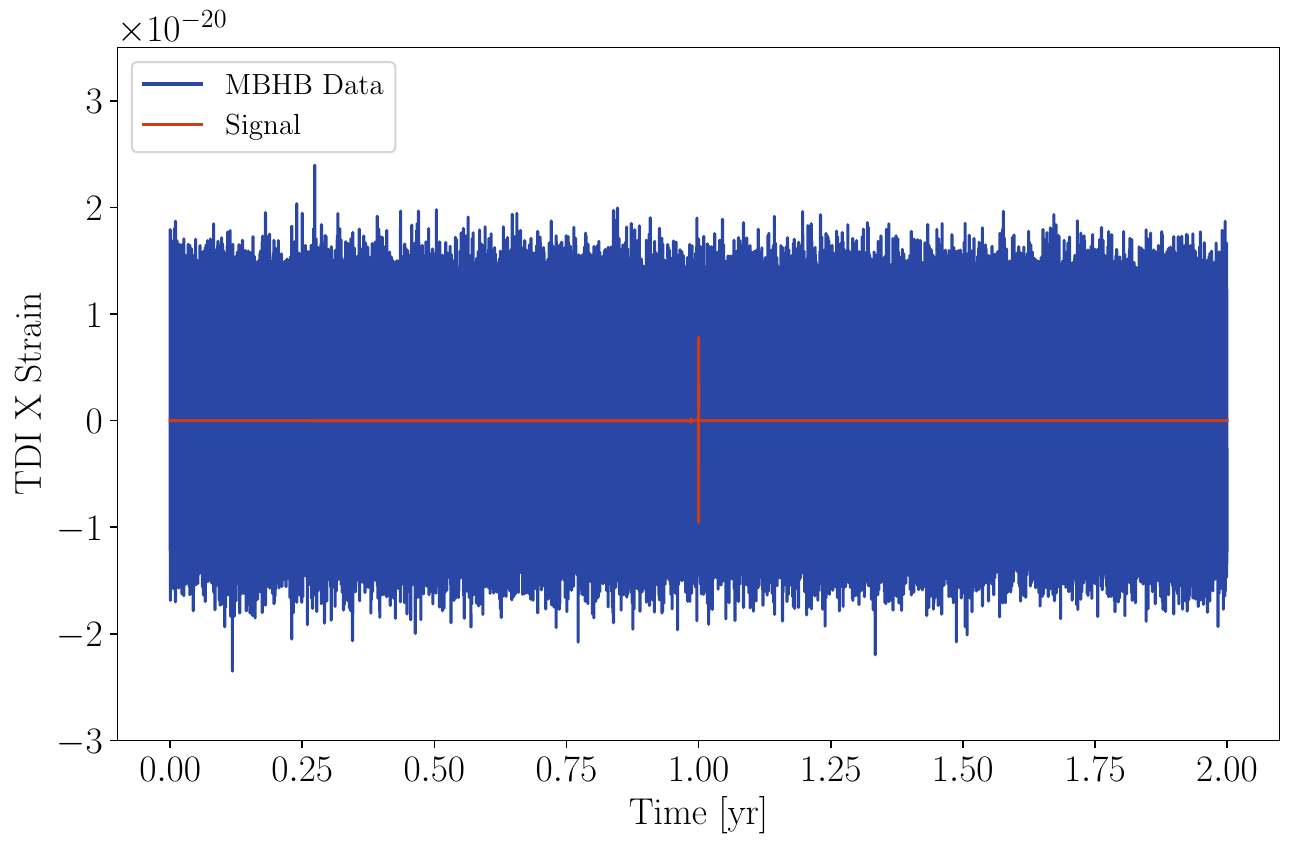}
			\caption{\label{fig:mbhb_data}}}
	\end{subfigure}\hfill
	\begin{subfigure}[t]{0.49\textwidth}{
			\includegraphics[width=\textwidth,valign=t]{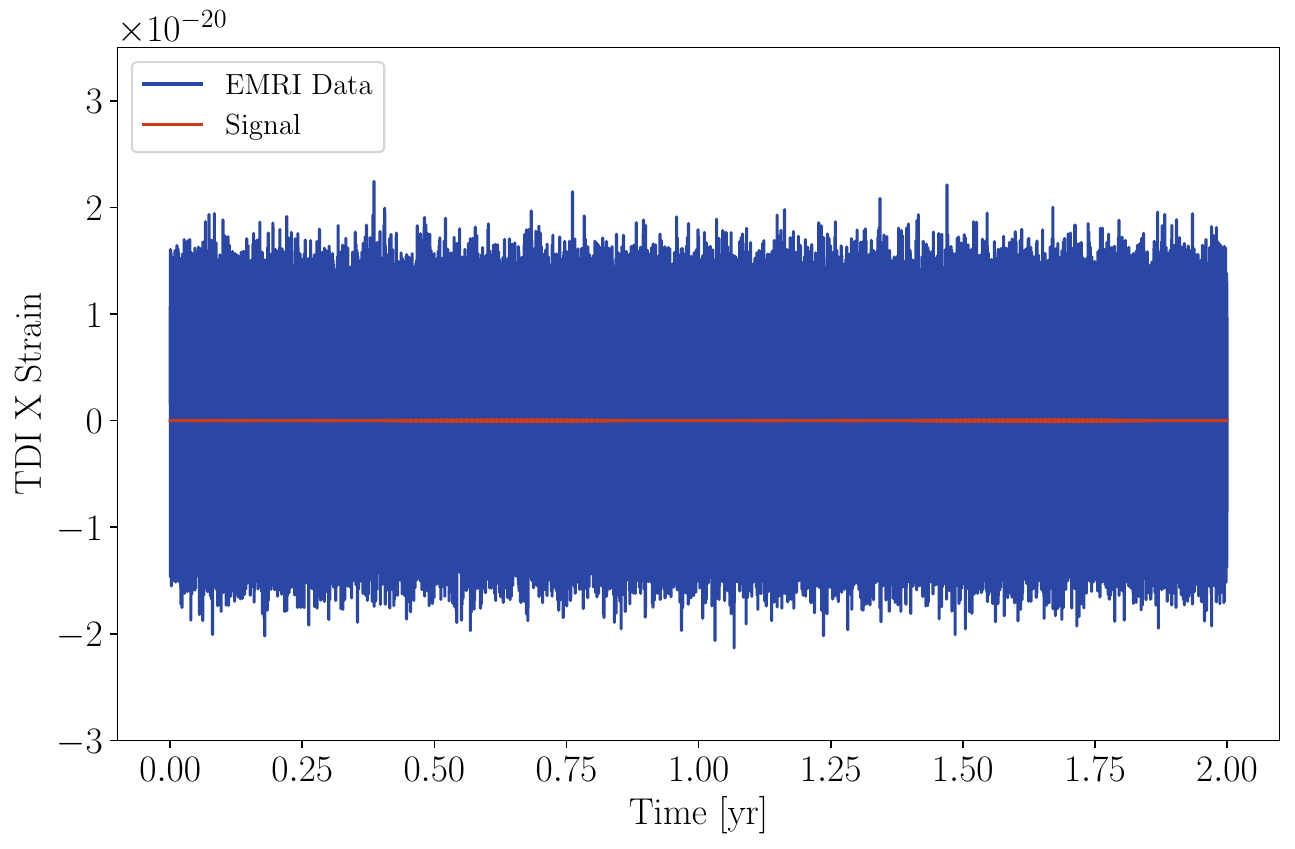}
			\caption{\label{fig:emri_data}}}
	\end{subfigure}
	\begin{subfigure}[t]{0.49\textwidth}{
			\includegraphics[width=\textwidth,valign=t]{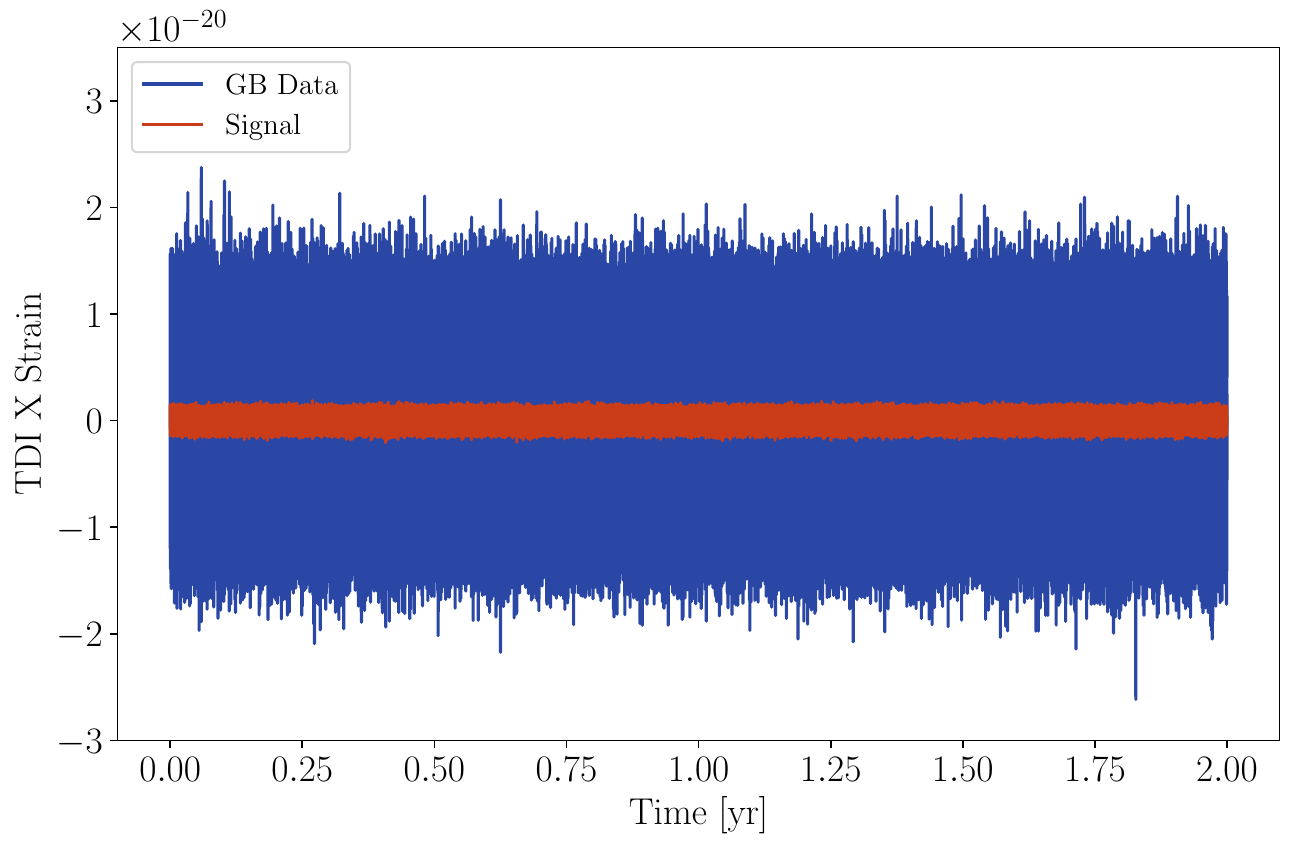}
			\caption{\label{fig:vgb_data}}}
	\end{subfigure}\hfill
	\begin{subfigure}[t]{0.49\textwidth}{
			\includegraphics[width=\textwidth,valign=t]{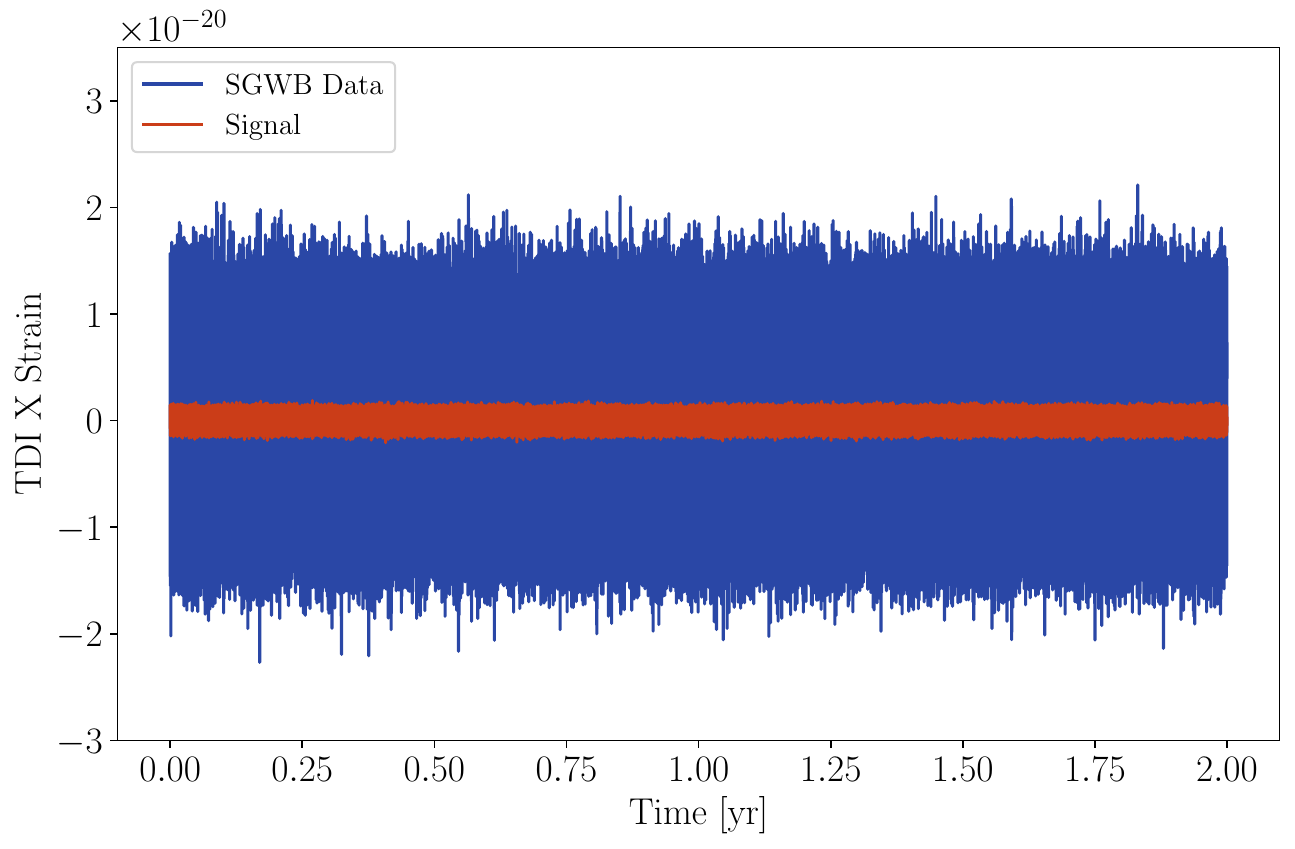}
			\caption{\label{fig:sgwb_data}}}
	\end{subfigure}
	\caption{\textbf{Showcase of various types of synthetic data generated:} This figure illustrates the composition of synthetic data, where the blue line represents the combined signal and noise, and the red line indicates the signal. Each sub-figure highlights a different source type: (a) MBHB, (b) EMRI, (c) GB, and (d) SGWB.}
	\label{fig:data_sample}
\end{figure}

% TDC

The data generation module within our \acl{GWDA} software platform is crucial, particularly for synthesizing data for training and validating machine learning models. This module meticulously simulates a spectrum of \ac{GW} scenario that encompasses a variety of astrophysical sources. It forms an integral part of the platform and can be integrated with other components. The data module is composed of \ac{GW} waveform, detector response and \ac{TDI} combination (see Fig. \ref{fig:data}).
Each of these aspects is critical in providing high-quality and synthetic data (see Fig. \ref{fig:data_sample}).
% and ensuring our platform's models are effectively trained to interpret real-world \ac{GW} observations with precision and accuracy.

\paragraph{GW Waveform}
In our data generation module, we have integrated a comprehensive range of \ac{GW} waveforms encompassing various astrophysical sources, including \ac{MBHB}, \ac{EMRI}, \ac{GB}, and the \ac{SGWB}. These waveforms represent the diverse ripples in spacetime caused by different cosmic phenomena \cite{amaro-seoane_laser_2017}. Each category possesses unique signal characteristics, reflecting distinct source properties like mass, spin, and orbital dynamics.
% Our module's capability to generate these varied waveform types is crucial for a holistic understanding of \ac{GW} sources.

\paragraph{Detector Response}
Our data generation module accounts for the orbital motion of spece-based \ac{GW} detectors, a crucial factor in accurately simulating \ac{GW} data. We have included advanced techniques to calculate the detector response using GPU acceleration, significantly enhancing the efficiency and speed of data processing \cite{katz_assessing_2022}. Additionally, our module offers a versatile API that supports arbitrary orbital trajectories, allowing for the precise modeling of a wide array of \ac{GW} detection scenarios \cite{ren_taiji_2023}.

\paragraph{TDI Combination}
\ac{TDI} is a technique particularly relevant for space-based \ac{GW} detectors. \ac{TDI} is used to combine data from multiple spacecraft in a constellation, compensating for the unequal arm lengths caused by their relative motion. This method effectively reduces laser frequency noise and enhances the detection sensitivity of the \ac{GW} signals \cite{otto_time-delay_2015}.

\subsubsection{Model}
\begin{figure}[ht!]
	\centering
	\includegraphics[width=0.8\textwidth]{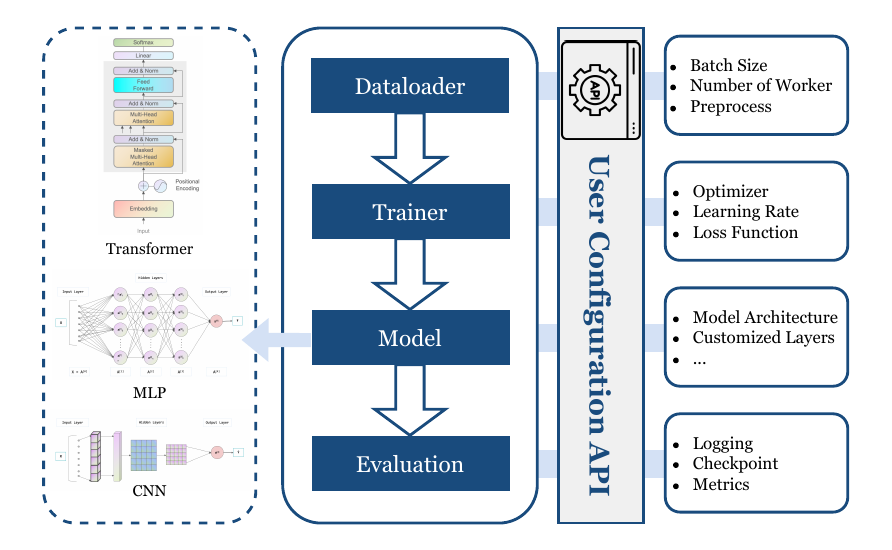}
	\caption{\textbf{Flowchart of AI model training and evaluation:} Presented here is the structured process for AI model training within our platform, illustrated on the left. It encompasses steps from data loading to performance benchmarking, culminating in the derivation of scientific results. On the right, the diagram emphasizes our platform's highly customizable configuration API, tailored to accommodate a diverse range of downstream tasks.}
	\label{fig:pipeline}
\end{figure}

In the realm of \acl{GWDA}, the implementation of advanced \ac{AI} models is pivotal for extracting and detecting complex signals from astrophysical sources. Our \acl{GWDA} platform harnesses a diverse array of basic models for further model construction. These basic models include \ac{MLP}, \ac{CNN}, and transformers.
% Each of them plays a crucial role in various \ac{AI} models used for \ac{GW} and analysis. 
Fig. \ref{fig:pipeline} shows how to train an \ac{AI} model within our GWAI platform.

\paragraph{MLP}
The \ac{MLP} forms a fundamental component of our \acl{GWDA} platform’s modeling toolkit.  \ac{MLP} is adept at pattern recognition and classification tasks, making it particularly useful for identifying the intricate patterns found in \ac{GW} data.

\paragraph{CNN}
\ac{CNN} \cite{lecun_gradient-based_1998} is used in our platform to leverage their superior capabilities in analyzing visual and time-series data. \acp{CNN} are particularly effective in handling the frequency and temporal aspects of \ac{GW} data, making them ideal for tasks such as signal detection or just feature extraction.

\paragraph{Transformer}
The Transformer model \cite{vaswani_attention_2017}, renowned for its success in \ac{NLP} \cite{devlin_bert_2019}, has been included in our platform for \acl{GWDA}. Its ability to handle sequential data makes it particularly effective for analyzing time-series data.

\subsubsection{Evaluation}

The comprehensive evaluation of our \acl{GWDA} platform is essential to validating its effectiveness and accuracy across various modules and tasks. This evaluation process is segmented into three key areas: data, model, and task. Each segment employs a range of benchmarking metrics and visualization techniques, crucial for assessing the platform's performance and reliability.

\paragraph{Data}
% 1. time domain waveform visualization
% 2. frequency domain waveform visualization
% 3. spectrograms
% 4. stationarity, Gaussianity test
% 5. match filtering SNR and usual SNR
In the data evaluation, we utilize a range of analytical tools and visualizations to assess the quality of the generated synthetic data.
% Time- and frequency-domain waveform visualizations offer insights into the basic properties of the signals, while spectrograms reveal the frequency content over time. 
The stationarity and gaussianity of the data are tested to ensure adherence to expected noise models. Additionally, \ac{SNR} measurements, both in terms of match filtering and usual SNR, are employed to evaluate the detectability and clarity of the \ac{GW} signals. These evaluations leverage widely available Python libraries, offering robust benchmarks of the data generated by our platform.

\paragraph{Model}
% 1. training/inference speed
% 2. JS/KL divergence as loss
% 3. activation map of CNN
% 4. attention map
% 5. common loss of DNN (MSE, MAE ...)
Model performance is critically assessed through various metrics. Training and inference speeds are measured to ensure the efficiency of our models. The Jensen-Shannon (JS) and Kullback-Leibler (KL) divergences are used as loss functions to quantify the similarity between the model outputs and target distributions. For \ac{CNN} models, activation maps are analyzed to understand feature extraction. For transformer models, attention maps are utilized to visualize focus areas in the data. Common losses like mean square error (MSE) and mean absolute error (MAE) are also tracked to gauge model accuracy.

\paragraph{Task}
% 1. Probability distribution visualization
% 2. ROC and AUC
% 3. False alarm rate of a GW event
% 4. P-P plot for bias visualization
% 5. overlap between output and target waveforms
Task-specific evaluations focus on the practical application and effectiveness of the models. Probability distribution visualizations aid in interpreting model outputs in a probabilistic framework. \ac{ROC} curves and \ac{AUC} metrics are employed to assess the classification performance. The false alarm rate of \ac{GW} events is a critical metric for detection reliability. P-P plots are used for visualizing biases in the model predictions, and the overlap between output and target waveforms provides a direct measure of model accuracy in waveform reconstruction.

\section{Illustrative examples}
This section presents the results obtained from various key tasks in \acl{GWDA} using our platform. Each subsection illustrates the effectiveness of the applied methods and models.

\begin{figure}[ht!]
	\centering
	\begin{subfigure}[t]{0.32\textwidth}{
			\includegraphics[width=\textwidth,valign=t]{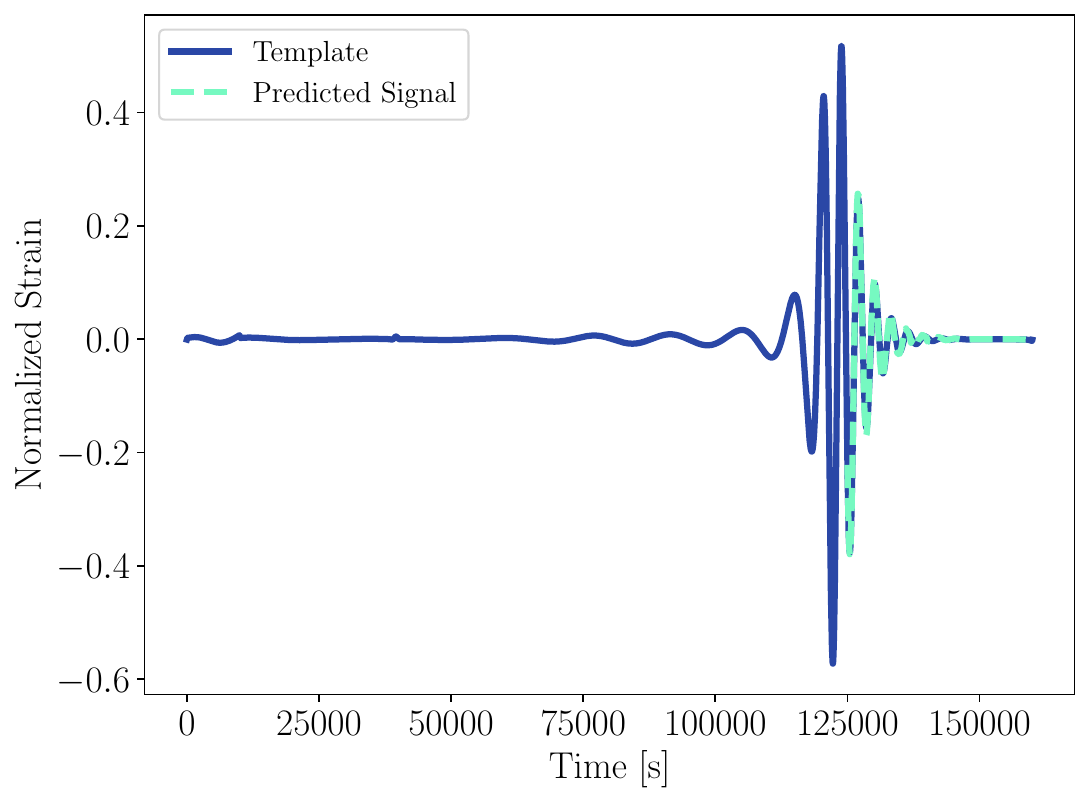}
			\caption{\label{fig:forecast}}}
	\end{subfigure}\hfill
	\begin{subfigure}[t]{0.333\textwidth}{
			\includegraphics[width=\textwidth,valign=t]{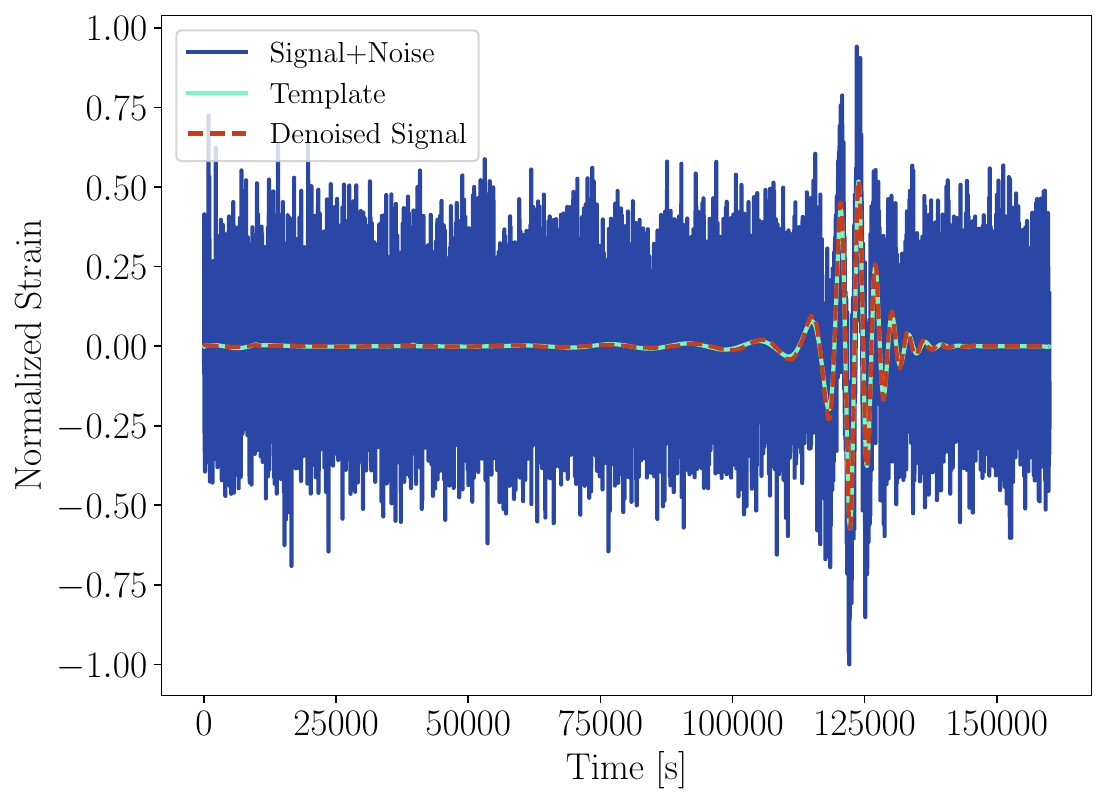}
			\caption{\label{fig:denoise}}}
	\end{subfigure}\hfill
	\begin{subfigure}[t]{0.34\textwidth}{
			\includegraphics[width=\textwidth,valign=t]{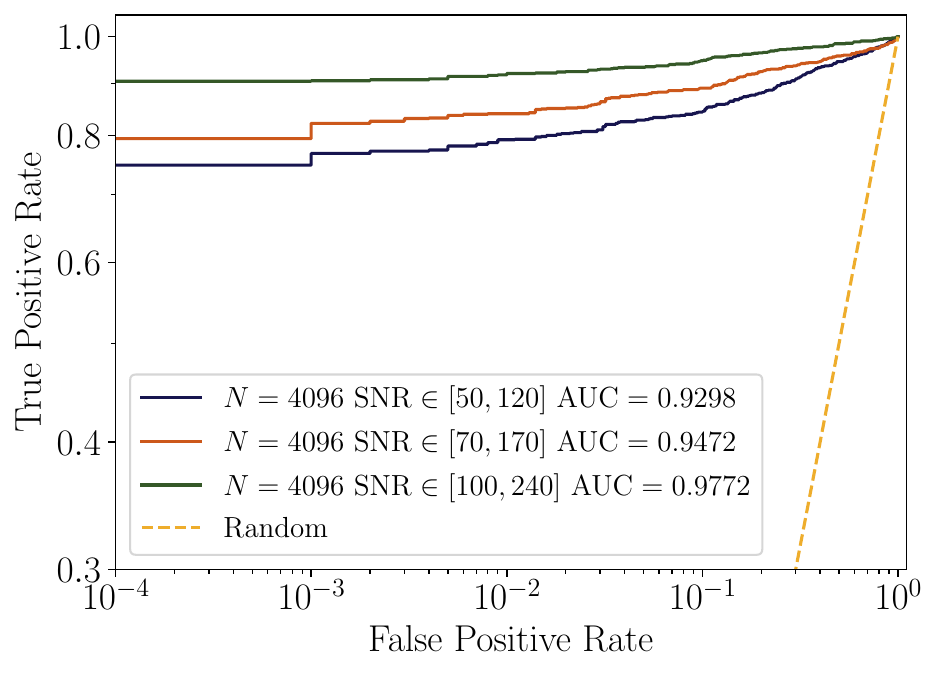}
			\caption{\label{fig:detection}}}
	\end{subfigure}
	\caption{\textbf{Results for typical downstream tasks:} This figure presents the outcomes of applying our models to key \acl{GWDA} tasks. (a) Waveform forecasting using CBS-GPT, where the blue line depicts the provided waveform template and the green line shows the predicted waveform. (b) \Ac{GW} denoising with an attention-based model, illustrated by a noisy waveform in blue, the original template in green, and the denoised waveform in orange. (c) The ROC curve for the DECODE model's performance across various samples, with $N$ representing the number of subsampling grid points. These results are adapted from the following sources: \cite{shi_compact_2024,zhao_space-based_2023,zhao_dilated_2024}.}
	\label{fig:result}
\end{figure}

\subsection{Waveform Forecasting}
% CBS-GPT (Fig. waveform)
Our platform employs the CBS-GPT \cite{shi_compact_2023} for waveform forecasting, achieving significant accuracy in predicting gravitational waveforms. Fig. \ref{fig:forecast} illustrates the forecasting results, showing a high degree of consistency between the predicted and actual waveforms.

\subsection{Denoising}
In the denoising task within the GWAI platform, we have integrated WaveFormer \cite{ren_intelligent_2022} and another advanced attention-based model \cite{zhao_space-based_2023}, both specifically tailored for denoising \ac{GW} ground-based and space-based \ac{GW} data. The attention-based model excels at extracting space-based \ac{GW} signals from the noisy background. The results, as illustrated in Fig. \ref{fig:denoise}, show a remarkable improvement in signal clarity. This advancement in denoising is particularly significant for space-based \acl{GWDA}, where data quality is essential for accurate signal detection.

\subsection{Detection}
For the detection task, we employed the DECODE model \cite{zhao_decode_2023}, which demonstrates exceptional proficiency in detecting \ac{EMRI} signals. The performance of DECODE is vividly illustrated in Fig. \ref{fig:detection}, through its \ac{ROC} curve. This model achieves a notably high \ac{AUC}, underscoring its precision in \ac{EMRI} detection. A key feature contributing to its success is the implementation of dilated convolution, which significantly enhances the model's capability to process the datasets, which span one year.

\section{Impact}
The field of \ac{GW} detection is rapidly advancing, with a particular focus on space-based detection initiatives. Notably, the \ac{ESA} has approved the \ac{LISA} project, marking a significant milestone in the quest to observe \acp{GW} from space. Additionally, other ambitious programs such as Taiji, TianQin, and \ac{DECIGO} are underway, each contributing to the global effort towards space-based \ac{GW} detection. These initiatives open the door to a plethora of potential scientific discoveries, offering new insights into the universe and furthering our understanding of astrophysical phenomena.
% The global scientific community is increasingly focused on these space-based detection projects, recognizing their potential to revolutionize our knowledge of the cosmos.

In recent years, the application of \ac{AI} in scientific research has seen remarkable growth, fundamentally transforming the research paradigm across various fields.
%\ac{AI} has demonstrated its profound capabilities in enhancing data analysis, interpretation, and prediction, leading to groundbreaking advancements. 
In the domain of \acl{GWDA}, especially in the context of space-based detection, leveraging \acg{AI} power is crucial for navigating the vast and complex datasets involved. However, harnessing \ac{AI} effectively requires robust software and hardware infrastructure. Our GWAI platform emerges as the first \ac{AI}-centered platform in \acl{GWDA}, designed to meet these challenges head-on.
%By integrating advanced \ac{AI} algorithms and computational techniques, GWAI sets a new standard for research in this field, facilitating the exploration of \acp{GW} with unprecedented efficiency and precision.
Looking forward, GWAI has the potential to advance gravitational wave data analysis by enhancing detection sensitivity and accuracy through advanced AI algorithms.
% Its flexible and modular architecture can adapt to incorporate new AI techniques, ensuring the platform remains at the forefront of technological advancements. 
By streamlining the data processing pipeline, GWAI can make the workflow more efficient and accessible, supporting collaborative research and data sharing. Ultimately, GWAI is poised to enable new discoveries and deepen our understanding of the universe in the rapidly evolving field of gravitational wave astronomy.

\section{Conclusions}
In this paper, we have introduced a groundbreaking AI-based \acl{GWDA} software platform, highlighting its comprehensive coverage, user-centric design, and advanced technological integration. Our platform stands out for its simplicity, modularity, flexibility, and ease of use, supported by extensive documentation and example showcases. The platform's successful implementation in various \acl{GWDA} scenarios demonstrates its effectiveness in enhancing \ac{GW} research. By integrating sophisticated \ac{AI} algorithms and offering a user-friendly experience, this platform makes significant strides in democratizing \acl{GWDA}, allowing researchers to focus on scientific discovery rather than technical complexities.

\section*{Declaration of Competing Interest}
The authors declare that they have no known competing financial interests or personal relationships that could have appeared to influence the work reported in this paper.

\section*{Acknowledgements}
The research was supported by the Peng Cheng Laboratory and by Peng Cheng Laboratory Cloud-Brain.
This work was also supported in part by the National Key Research and Development Program of China Grant No.~2021YFC2203001 and in part by the NSFC (No.~11920101003 and No.~12021003). Z.C was supported by the ``Interdisciplinary Research Funds of Beijing Normal University'' and CAS Project for Young Scientists in Basic Research YSBR-006.

\bibliographystyle{apsrev4-2-1}
\bibliography{ref}

\appendix
\section{Ablation Analysis on Model Architecture}

\begin{table}[h!]
	\centering
	\begin{threeparttable}
		\caption{\textbf{Ablation Study on the WaveFormer Model} This table shows the results of an ablation study on the WaveFormer model, evaluating denoising performance on three LIGO O1 events (GW150914, GW151012, GW151226) using overlap as the performance metric. The model's components include One-Class Embedding (OCE), Gaussian-weighted Self-Attention (GSA), relative positional encoding (Rotray), and residual connections. The highest overlap for each event is highlighted in bold. Results demonstrate how different combinations of these components impact model performance.}
		\label{tab:ablation}
		\begin{tabular}{@{}ccccccc@{}}
			\toprule
			\textbf{OCE}   & \textbf{GSA}   & \textbf{Rotray} & \textbf{Residual} & \textbf{GW150914} & \textbf{GW151012} & \textbf{GW151226} \\ \midrule
			\XSolidBrush   & \CheckmarkBold & \XSolidBrush    & \XSolidBrush      & 0.900             & 0.730             & 0.470             \\
			\XSolidBrush   & \XSolidBrush   & \XSolidBrush    & \XSolidBrush      & 0.900             & 0.720             & 0.810             \\
			\CheckmarkBold & \XSolidBrush   & V1              & \CheckmarkBold    & 0.884             & 0.841             & 0.852             \\
			\CheckmarkBold & \XSolidBrush   & V2              & \CheckmarkBold    & 0.878             & 0.754             & 0.875             \\
			\XSolidBrush   & \XSolidBrush   & \XSolidBrush    & \CheckmarkBold    & 0.913             & 0.783             & 0.902             \\
			\CheckmarkBold & \XSolidBrush   & \XSolidBrush    & \CheckmarkBold    & \textbf{0.977}    & \textbf{0.944}    & \textbf{0.959}    \\
			\bottomrule
		\end{tabular}
	\end{threeparttable}
\end{table}

% \textcolor{myred}{
The ablation study results in Table \ref{tab:ablation} reveal the impact of different components of the WaveFormer model on its denoising performance for three LIGO O1 events (GW150914, GW151012, and GW151226). The highest overall performance is achieved when both One-Class Embedding (OCE) and residual connections are enabled, with the model delivering significant improvements in overlap scores, especially for GW150914 (0.977) and GW151226 (0.959). Interestingly, the use of Gaussian-weighted Self-Attention (GSA) without OCE consistently underperforms, highlighting the importance of OCE for effective signal processing. Additionally, relative positional encoding (Rotray) shows moderate influence on performance, with V2 slightly outperforming V1. These findings underscore the critical role of OCE and residual connections in enhancing the model’s capacity to accurately denoise gravitational wave signals, while components like GSA and positional encoding appear to be less impactful on their own.
% }

\section{Hyperparameter Sensitivity Analysis}

\begin{table}[h]
	\centering
	\begin{threeparttable}
		\caption{\textbf{Hyperparameter Sensitivity of the  Space-Based GW Denoising Model} This table presents the results of the ablation study conducted on the denoising model, specifically analyzing the impact of Long-Time Transformer Blocks (LTTB) and Short-Time Transformer Blocks (STTB) on performance metrics. The SI-SNR (Scale-Invariant Signal-to-Noise Ratio) loss function is used specifically for denoising tasks \cite{vincent_performance_2006}.}
		\label{tab:hyper}
		\begin{tabular}{@{}ccll@{}}
			\toprule
			\textbf{\# of LTTB} & \textbf{\# of STTB} & \textbf{Data Length} & \textbf{SI-SNR $(\uparrow)$} \\ \midrule
			2                   & 2                   & 4,000                & 14.02 dB                     \\
			2                   & 2                   & 8,000                & 15.48 dB                     \\
			2                   & 2                   & 8,000                & 14.24 dB (warm up)           \\
			2                   & 2                   & 16,000               & 16.32 dB                     \\
			4                   & 4                   & 16,000               & \textbf{19.90 dB}            \\ \bottomrule
		\end{tabular}
		\begin{tablenotes}
			\small
			\item \textbf{Note:} The input data SNR is approximately -35 dB.
		\end{tablenotes}
	\end{threeparttable}
\end{table}

% \textcolor{myred}{
We further conducted tests under different hyper-parameters, as detailed in Tab. \ref{tab:hyper}, and found that the model significantly benefits from its ability to model long-range dependencies. Specifically, providing longer input data results in improved denoising performance. However, our analysis also revealed that the warm-up strategy does not enhance performance in this context, suggesting it may not be suitable for this specific denoising task. For further details about the model architecture and configuration, please refer to \cite{zhao_space-based_2023}.
% }

\begin{figure}[ht!]
	\centering
	\begin{subfigure}[t]{0.49\textwidth}{
			\includegraphics[width=\textwidth,valign=t]{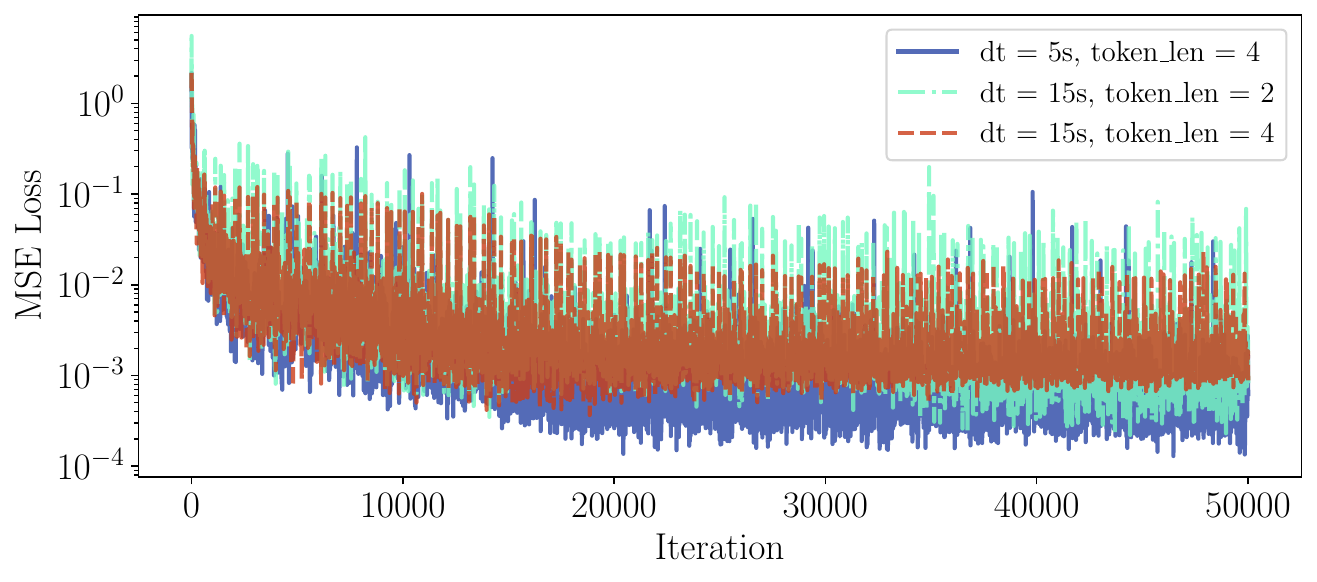}
			\caption{\label{fig:emri_loss}}}
	\end{subfigure}\hfill
	\begin{subfigure}[t]{0.49\textwidth}{
			\includegraphics[width=\textwidth,valign=t]{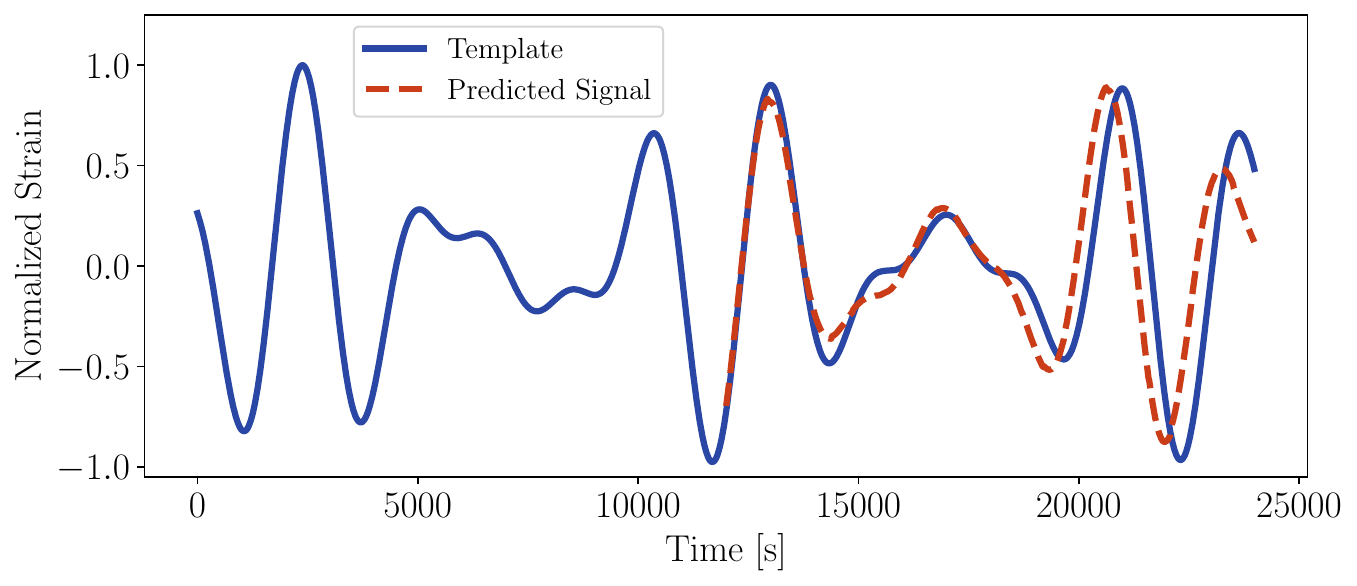}
			\caption{\label{fig:emri_forecast}}}
	\end{subfigure}
	\caption{\textbf{Hyperparameter Sensitivity Analysis of CBS-GPT.} This figure illustrates the impact of different hyperparameter settings on the performance of the CBS-GPT model. (a) Shows the validation loss across different sampling rates (5s, 15s) and token lengths (2, 4), with the model using a sampling rate of 5s and a token length of 4 achieving the best performance. (b) Displays a showcase of the output generated by the best-performing model identified in (a).}
	\label{fig:hyper_loss}
\end{figure}

% \textcolor{myred}{
The hyperparameter sensitivity analysis of CBS-GPT, as shown in Fig. \ref{fig:emri_loss}, demonstrates the significant impact of sampling rate and token length (the number of sampling points grouped into a single token) on model performance. The model with a sampling rate of 5 seconds and a token length of 4 achieves the lowest validation loss, suggesting that this configuration strikes an optimal balance between temporal resolution and data representation within each token for effective learning. In Fig. \ref{fig:emri_forecast}, despite the inherent challenge of modeling EMRI due to their complex and eccentric waveforms, our model successfully captures the waveform dynamics, highlighting the robustness of this configuration. These findings emphasize the importance of carefully tuning hyperparameters such as sampling rate and token length to maximize the performance of CBS-GPT in gravitational wave data analysis.
% }

\end{document}